\documentstyle[preprint,aps]{revtex}

\input epsf         
\epsfverbosetrue    
\def\postscript#1{\begin{center}\leavevmode
\hbox{\epsfxsize=1.0\columnwidth\epsfbox{#1}}\end{center}}

\begin{document}

\draft

\title{Quantized hydrodynamic model and
the dynamic structure factor for a trapped Bose gas}

\author{Wen-Chin Wu and A. Griffin}
\address{Department of Physics, University of Toronto,
         Toronto, Ontario, Canada, M5S 1A7} 

\date{submitted to Phys. Rev. A on June 7, 1996}

\maketitle

\begin{abstract} 
We quantize the recent hydrodynamic analysis of Stringari for the low-energy 
collective modes of a trapped Bose gas at $T=0$. This is based on the 
time-dependent Gross-Pitaevskii equation, but omits the kinetic energy of 
the density fluctuations. We diagonalize the hydrodynamic Hamiltonian 
in terms of the normal modes associated with the amplitude and phase 
of the inhomogeneous Bose order parameter. These normal modes provide a 
convenient basis for calculating observable quantities.  As applications, we 
calculate the depletion of the condensate at $T=0$ as well as the inelastic 
light-scattering cross section $S({\bf q},\omega)$ from low-energy condensate 
fluctuations. The latter involves a sum over all normal modes, with a weight 
proportional to the square of the ${\bf q}$ Fourier component of the density 
fluctuation associated with a given mode. Finally, we show how the 
Thomas-Fermi hydrodynamic description can be derived starting from the 
coupled Bogoliubov equations.
\end{abstract}

\vskip 0.1 true in


\narrowtext

\section{INTRODUCTION}

The recent achievement of Bose-Einstein condensation (BEC)
in trapped atomic gases at ultra-cold temperatures
\cite{AEMWC95} has stimulated interest in the collective oscillations of 
a non-uniform Bose condensate \cite{ERBC96,Fetter96-1}. 
These are of direct experimental interest \cite{ERBDC96}.
At temperatures well below the BEC 
transition temperature, all atoms are in the condensate, and one
may then base the analysis on the Gross-Pitaevskii (GP) equation of motion
\cite{Gross61} for the time-dependent condensate wave function
$\Phi({\bf r},t)$ \cite{NP90}.
Linearizing around the static equilibrium value
$\Phi_0({\bf r})$, this leads to the well-known Bogoliubov
equations of motion for the excitations of the condensate 
(see, for example, Refs.~\cite{Fetter72,Griffin96}).
Recently Stringari \cite{Stringari96-2} has noted that
the GP equation can be rewritten as equations for the condensate density
and phase variables and that within a Thomas-Fermi approximation (TFA)
which neglects the kinetic energy associated with density fluctuations, 
these have the familiar structure of the hydrodynamic equations of a 
superfluid at $T=0$ \cite{NP90}. This allows one to give a simple 
theory of the low-frequency excitations of a trapped Bose gas,
without the necessity of solving the full set of coupled Bogoliubov equations
\cite{ERBC96,Fetter96-1}.

In the present paper, we develop the 
approach of Ref.~\cite{Stringari96-2} by quantizing this ``hydrodynamic''
description and diagonalizing the associated Hamiltonian in terms of the 
normal modes of the condensate. This is a natural
generalization of the well-known discussion used in a {\em uniform}
Bose-condensed
fluid \cite{LP80}. These modes form a natural basis for
understanding the effect of low-energy condensate fluctuations on
various physical quantities.
We illustrate this in this paper
by evaluating the local depletion of the condensate at 
$T=0$ as well as the inelastic light
scattering cross section from the low-energy
condensate fluctuations. The latter is found to be quite different from the
case of a {\em uniform}
weakly-interacting Bose gas (as discussed, for example,
in Refs.~\cite{Griffin93,GW96}). 

The present ``hydrodynamic'' description is only 
an adequate description of the low-energy excitations of the
condensate at $T=0$. In particular, as discussed by Stringari 
\cite{Stringari96-2}, the neglect of the kinetic energy in the TF
approximation is only correct in the strong interaction
or large-density limit \cite{BP96}. 
In Section~V, we discuss the precise relation
between the present analysis and that based on the full coupled Bogoliubov
equations for condensate excitations. 
In spite of its deficiencies, 
the hydrodynamic approximation has the advantage that it allows
one to illustrate the qualitative effects of the low-frequency modes 
in a very explicit manner.

\section{Quantized Hydrodynamic Theory}

In a dilute trapped Bose gas \cite{AEMWC95}
near $T=0$, most of the atoms are Bose-condensed and 
thus the whole system can be well described by the Gross-Pitaevskii 
Hamiltonian \cite{Gross61,NP90}
\begin{eqnarray}
H = \int d{\bf r}\ \Phi^*({\bf r})
\left[-\frac{\hbar^2\nabla^2}{2m} - \mu+V_{\rm ex}({\bf r}) +
{1\over 2}g_0 |\Phi({\bf r})|^2 \right]\ \Phi ({\bf r}).
\label{eq:H}
\end{eqnarray}
In (\ref{eq:H}), the static external potential $V_{\rm ex}({\bf r})$ 
is responsible for trapping of the atoms.
We have assumed a short-range interaction
$g_0\equiv 4\pi \hbar^2 a/m$ between the atoms, with $a$ being the 
{\it s}-wave scattering length. 
The Bose-condensate order parameter $\Phi({\bf r})$ can also be written as

\begin{equation}
\Phi ({\bf r})=|\Phi ({\bf r})|e^{i\phi ({\bf r})},
\label{eq:o.p.}
\end{equation}
in terms of amplitude and phase components.
Substitution of (\ref{eq:o.p.}) into (\ref{eq:H}) yields 
\begin{eqnarray}
H = \int d{\bf r}\ \left[ \sqrt{\rho_c}\left(-{\hbar^2\over 2m}\right)
\nabla^2\sqrt{\rho_c}+{1\over 2}m\rho_c {\bf v}^2-\rho_c\mu+\rho_c V_{\rm ex}
+{1\over 2}g_0\rho_c^2 \right],
\label{eq:H.1}
\end{eqnarray}
where the local condensate density and superfluid velocity are defined by

\begin{mathletters}
\label{eq:rho-v}
\begin{eqnarray}
\rho_c({\bf r})&\equiv&|\Phi ({\bf r})|^2,\label{eq:rho-1}\\
{\bf v}({\bf r})&\equiv&{\hbar\over 2mi|\Phi ({\bf r})|^2}
\left[\Phi^*({\bf r})\nabla\Phi ({\bf r})-\nabla\Phi^*({\bf r})
\Phi ({\bf r}) \right]\nonumber\\
&=&{\hbar\nabla\phi ({\bf r})\over m}.
\label{eq:v-1}
\end{eqnarray}
{\rm Expressed in terms of these variables, the GP
theory can be formulated in terms of
hydrodynamic-type equations.}
\end{mathletters}
We can also consider a time-dependent order parameter
$\Phi({\bf r},t)$ given by the time-dependent GP equation \cite{NP90}

\begin{eqnarray}
i\hbar{\partial\over \partial t}\Phi ({\bf r},t)=
\left[-\frac{\hbar^2\nabla^2}{2m} - \mu+V_{\rm ex}({\bf r}) +
g_0 |\Phi({\bf r},t)|^2 \right]\ \Phi ({\bf r},t),
\label{eq:H.GP}
\end{eqnarray}
which can be used to discuss the excitations from the equilibrium 
condensate wavefunction $\Phi_0({\bf r})$. 

To quantize a hydrodynamic-type description, we first replace
the dynamical variables in (\ref{eq:rho-v})
by the corresponding quantum mechanical operators.
The density and velocity operators 
satisfy the exact commutation rule \cite{LP80}
\begin{equation}
[\hat{\rho}({\bf r}), \hat{\bf v}({\bf r^\prime})]
=i{\hbar\over m}\nabla\delta ({\bf r-r^\prime}).
\label{eq:commutation.1}
\end{equation}
This can be used in the analysis of the $T=0\ $ GP expression
(\ref{eq:rho-v}) because almost all the atoms are in the condensate
and hence $\hat{\rho}_c({\bf r})\simeq\hat{\rho}({\bf r})$. 
(By the same token, it is difficult to quantize a hydrodynamic-type
theory at $T\neq 0$ when there is significant depletion.)
In terms of the phase operator defined by (\ref{eq:o.p.}),
(\ref{eq:commutation.1}) is equivalent to \cite{LP80}
\begin{equation}
[\hat{\rho}({\bf r}), \hat\phi({\bf r^\prime})]=
i\delta({\bf r-r^\prime}).
\label{eq:commutation.2}
\end{equation}
Using the operator version of the Hamiltonian in
(\ref{eq:H.1}), it is straightforward to
work out the Heisenberg equations of motion for both
$\hat{\rho}$ and $\hat{\bf v}$ making use of (\ref{eq:commutation.1}). 
One obtains \cite{Stringari96-2}

\begin{equation}
{\partial\hat{\rho}({\bf r},t)\over \partial t}=
-\nabla\cdot\left[\hat{\rho}({\bf r},t)\hat{\bf v}({\bf r},t)\right]
\label{eq:key.1}
\end{equation}
and
\begin{equation}
m{\partial\hat{\bf v}({\bf r},t)\over \partial t}=
-\nabla\left[V_{\rm ex}({\bf r})-\mu +g_0\hat{\rho}({\bf r},t)
-{\hbar^2\nabla^2\sqrt{\hat{\rho}({\bf r},t)}\over 
2m\sqrt{\hat{\rho}({\bf r},t)}}+{1\over 2}m\hat{\bf v}^2({\bf r},t)\right].
\label{eq:key.2}
\end{equation}
Assuming the validity of the Gross-Pitaevskii \cite{Gross61} Hamiltonian
given by (\ref{eq:H}), (\ref{eq:key.1}) and 
(\ref{eq:key.2}) are exact and can be used to discuss 
the dynamics of the condensate. They are equivalent to the GP
equation in (\ref{eq:H.GP}).

It is useful to separate various operators
into equilibrium and small fluctuation parts
\begin{eqnarray}
\hat{\Phi}({\bf r},t)&\equiv& \Phi_0({\bf r})+\delta\hat{\Phi}({\bf r},t)
\nonumber\\
\hat{\rho}({\bf r},t)&\equiv& \rho_0({\bf r})+\delta\hat{\rho}({\bf r},t)
\nonumber\\
\hat{\phi}({\bf r},t)&\equiv& \phi_0({\bf r})+\delta\hat{\phi}({\bf r},t)
\nonumber\\
\hat{\bf v}({\bf r},t)&\equiv&{\bf v}_0({\bf r})+\delta\hat{\bf v}({\bf r},t),
\label{eq:rho}
\end{eqnarray}
where the static condensate density profile is 
$\rho_0({\bf r})\equiv |\Phi_0({\bf r})|^2$ and 
${\bf v}_0({\bf r})\equiv \hbar\nabla\phi_0({\bf r})/m$.
Before preceeding, we recall that the equilibrium values are given by the 
solutions of the time-independent GP equation [see (\ref{eq:H.GP})],

\begin{equation}
\left[-\frac{\hbar^2\nabla^2}{2m} - \mu+V_{\rm ex}({\bf r}) +
g_0 \rho_0({\bf r}) \right]\ \Phi_0 ({\bf r})=0,
\label{eq:GP_static}
\end{equation}
where $\Phi_0({\bf r})=|\Phi_0({\bf r})|e^{i\phi_0({\bf r})}$.
The static solution of (\ref{eq:key.1}) must satisfy

\begin{equation}
\nabla\cdot\left[|\Phi_0({\bf r})|^2{\bf v}_0({\bf r})\right]=0,
\label{eq:key.1_static}
\end{equation}
which is equivalent to

\begin{equation}
2\nabla|\Phi_0({\bf r})|\cdot\nabla\phi_0({\bf r})+
|\Phi_0({\bf r})|\nabla^2\phi_0({\bf r})=0.
\label{eq:key.1_static_equivalent}
\end{equation}
Using this in (\ref{eq:GP_static}), the latter equation can be 
reduced to an equation for the amplitude
$|\Phi_0({\bf r})|=\sqrt{\rho_0({\bf r})}$ 

\begin{equation}
\left[-\frac{\hbar^2\nabla^2}{2m} - \mu+V_{\rm ex}({\bf r}) +
g_0 \rho_0({\bf r}) +{1\over 2}m{\bf v}_0^2({\bf r})\right]\ 
|\Phi_0 ({\bf r})|=0.
\label{eq:key.2_static}
\end{equation}
Solutions of (\ref{eq:key.1_static_equivalent}) and 
(\ref{eq:key.2_static}) with a spatially-varying phase $\phi_0({\bf r})$
correspond to states with a non-zero superfluid velocity
${\bf v}_0({\bf r})$ (such as vortices).

In this paper, we are interested in small fluctuations of the static 
condensate $\Phi_0({\bf r})$.  In terms of (\ref{eq:rho-1}) and
(\ref{eq:rho}), we obtain
\begin{equation}
\delta\hat{\rho}({\bf r},t)=\Phi^*_0({\bf r})
\delta\hat{\Phi}({\bf r},t)+\Phi_0({\bf r})
\delta\hat{\Phi}^\dagger({\bf r},t). 
\label{eq:delta_rho}
\end{equation}
Writing $\hat{\Phi}({\bf r},t)\equiv \Phi_0({\bf r})
[1+\delta\hat{A}({\bf r},t)]e^{i\delta\hat{\phi}({\bf r},t)}$ 
in terms of Hermitian amplitude and phase fluctuation operators, we find
\begin{equation}
\delta\hat{\Phi}({\bf r},t)=\Phi_0({\bf r})[\delta\hat{A}({\bf r},t)
+i\delta\hat{\phi}({\bf r},t)]
\label{eq:delta_Phi}
\end{equation}
and hence the density fluctuation operator in (\ref{eq:delta_rho}) is 
related (as expected) to the amplitude fluctuation operator

\begin{equation}
\delta\hat{\rho}({\bf r},t)=2|\Phi_0({\bf r})|^2\delta\hat{A}({\bf r},t)
=2\rho_0({\bf r})\delta\hat{A}({\bf r},t).
\label{eq:delta_rho.1}
\end{equation}
We have used the fact that $\delta\hat{A}=\delta\hat{A}^\dagger$
and $\delta\hat{\phi}=\delta\hat{\phi}^\dagger$.
With (\ref{eq:rho}), the commutator in  
(\ref{eq:commutation.2}) can be rewritten in terms of fluctuation operators
\begin{equation}
[\delta\hat{\rho}({\bf r}), \delta\hat\phi({\bf r^\prime})]=
i\delta({\bf r-r^\prime}).
\label{eq:commutation.3}
\end{equation}

The key simplification to be used in analyzing the
Eqs.~(\ref{eq:key.1}) and (\ref{eq:key.2}) is to neglect the kinetic 
energy term associated with density fluctuations
$\hbar^2\nabla^2\sqrt{\hat{\rho}}/2m\sqrt{\hat{\rho}}$
in determining both the equilibrium and time-dependent solutions.
This is referred to as the Thomas-Fermi approximation (TFA)
\cite{BP96,Stringari96-2}. 
In the TFA, for ${\bf v}_0({\bf r})=0$, 
the static density profile $\rho_0({\bf r})$
is related to the external trapping potential by
\cite{statement1_paper6}

\begin{equation}
\mu=g_0\rho_0({\bf r})+V_{\rm ex}({\bf r}),
\label{eq:rho_0}
\end{equation}
where the chemical potential $\mu$ is fixed by the total number of atoms
$N=\int d{\bf r}\rho_0({\bf r})$ and the condition $\rho_0({\bf r})\geq 0$.
This should be a good description if the number
of atoms is large enough. 
More specifically, it is valid for \cite{BP96} $Na/a_{HO}\gg 1$,
where $a_{HO}\equiv(\hbar/m\omega_0)^{1\over 2}$ is the
characteristic harmonic oscillator length for an external potential 
$V_{\rm ex}({\bf r})={1\over 2}m\omega_0^2 r^2$. One can show that this
implies $\mu\gg \hbar\omega_0$.  The use of the TFA
precludes the consideration of a vortex
state, where $v_0({\bf r})=n\hbar/mr_\perp$
($r_\perp$ is the distance from the axis of the vortex
and $n$ is an integer).  In the presence of vortices,
the kinetic energy terms in (\ref{eq:key.2_static})
plays a crucial role in the determination of the spatial
variation of $|\Phi_0({\bf r})|$ \cite{DPS96}.

In the same approximation, (\ref{eq:key.1}) and (\ref{eq:key.2})
give (to first order in the perturbed quantities)

\begin{equation}
{\partial\delta\hat{\rho}({\bf r},t)\over \partial t}=
-\nabla\cdot\left[{\rho}_0({\bf r})\delta\hat{\bf v}({\bf r},t)\right]
\label{eq:rho.1}
\end{equation}
and
\begin{equation}
m{\partial\delta\hat{\bf v}({\bf r},t)\over \partial t}=
-\nabla\left[g_0\delta\hat{\rho}({\bf r},t)\right].
\label{eq:v.1}
\end{equation}
To be explicit, we have neglected the following
terms in the square bracket in (\ref{eq:key.2})

\begin{equation}
{\hbar^2\over 4m\sqrt{{\rho}_0({\bf r})}}\left\{
-{\nabla^2\sqrt{{\rho}_0({\bf r})}\over {\rho}_0({\bf r})}
\delta\hat{\rho}({\bf r},t)+\nabla^2\left[{
\delta\hat{\rho}({\bf r},t)\over \sqrt{{\rho}_0({\bf r})}}\right]\right\}
\label{eq:neglect}
\end{equation}
relative to $g_0\delta\hat{\rho}({\bf r},t)$.
Combining (\ref{eq:rho.1}) and (\ref{eq:v.1}) gives

\begin{equation}
{\partial^2\delta\hat{\rho}({\bf r},t)\over \partial t^2}=
\nabla\cdot\left[{g_0\over m}{\rho}_0({\bf r})\nabla\delta\hat{\rho}({\bf r},t)
\right]
\label{eq:rho-v.diff}
\end{equation}
and the equivalent
\begin{equation}
{\partial^2\delta\hat{\phi}({\bf r},t)\over \partial t^2}=
\nabla\cdot\left[{g_0\over m}{\rho}_0({\bf r})\nabla\delta\hat{\phi}({\bf r},t)
\right].
\label{eq:rho-v.diff_1}
\end{equation}
The solutions of (\ref{eq:rho-v.diff}) or (\ref{eq:rho-v.diff_1})
give the low-frequency condensate collective modes \cite{Stringari96-2}
of an inhomogeneous Bose gas with a local condensate density $\rho_0({\bf r})$
given by (\ref{eq:rho_0}).

In this approximate treatment, the GP Hamiltonian (\ref{eq:H.1}) reduces to 

\begin{equation}
\hat{H} = H_0+{1\over 2}\int d{\bf r}\ \left[m\rho_0({\bf r}) \delta
\hat{\bf v}^2({\bf r})+g_0\delta\hat{\rho}^2({\bf r}) \right],
\label{eq:H.2}
\end{equation}
where the ground state is described by

\begin{equation}
H_0\equiv -{1\over 2}\int \ d{\bf r} g_0\rho_0^2({\bf r}).
\label{eq:H_0}
\end{equation}
An expression identical to (\ref{eq:H.2}) is derived in
\S~5.2 of Ref.~\cite{NP90} by a different procedure.
Clearly the quadratic form (\ref{eq:H.2}) can be diagonalized
by expanding the operators $\delta\hat{\rho}({\bf r})$ and
$\delta\hat{\phi}({\bf r})$ [or $\delta\hat{\bf v}({\bf r})$]

\begin{eqnarray}
\delta\hat{\rho}({\bf r})&=&
{\sum_{j}} \left[A_j \psi_j({\bf r})\hat{\alpha}_j+
A^*_j \psi^*_j({\bf r})\hat{\alpha}^\dagger_j\right]
\nonumber\\
\delta\hat{\phi}({\bf r})&=&
{\sum_{j}}  \left[B_j \psi_j({\bf r})\hat{\alpha}_j+
B^*_j \psi^*_j({\bf r})\hat{\alpha}^\dagger_j\right],
\label{eq:quantized}
\end{eqnarray}
in terms of creation and annihilation operators $\hat{\alpha}^\dagger_j$
and $\hat{\alpha}_j$ which create and destroy excitations of the condensate
with energy $\hbar\omega_j$.
These operators satisfy the usual Bose commutation relations:
$[\hat{\alpha}_j, \hat{\alpha}_{j^\prime}]=
[\hat{\alpha}^\dagger_j, \hat{\alpha}^\dagger_{j^\prime}]=0$
and $[\hat{\alpha}_j, \hat{\alpha}^\dagger_{j^\prime}]=\delta_{j,j^\prime}$. 
The hermiticity of $\delta\hat{\rho}$ and $\delta\hat{\phi}$ is
ensured by the form used in (\ref{eq:quantized}).
The eigenfunctions $\psi_j({\bf r})$ are assumed to satisfy
both the normalization relation

\begin{equation}
\int d{\bf r}\ \psi_j^*({\bf r})\psi_{j^\prime}({\bf r})=\delta_{j,j^\prime},
\label{eq:normalization}
\end{equation}
and the completeness relation

\begin{equation}
\sum_{j} \psi_j^*({\bf r})\psi_j({\bf r}^\prime)=\delta({\bf r-r^\prime}).
\label{eq:completeness}
\end{equation}
Using (\ref{eq:quantized}) with the help of
(\ref{eq:completeness}), the commutator
(\ref{eq:commutation.3}) can be shown to require that $A^*_j B_j=-i/2$. 

Assuming that (\ref{eq:H.2}) has been diagonalized to give

\begin{equation}
\hat{H}=H_0+{\sum_{j}}\
\hbar \omega_j \left(\hat{\alpha}^\dagger_j\hat{\alpha}_j
+{1\over 2}\right),
\label{eq:H.3}
\end{equation}
the time-dependent operators in Heisenberg representation are 
then immediately given by

\begin{eqnarray}
\delta\hat{\rho}({\bf r},t)&=&
{\sum_{j}} \left[A_j \psi_j({\bf r})e^{-i\omega_j t}\hat{\alpha}_j+
A^*_j \psi^*_j({\bf r})e^{i\omega_j t}\hat{\alpha}^\dagger_j\right]
\nonumber\\
\delta\hat{\phi}({\bf r},t)&=&
{\sum_{j}}  \left[B_j \psi_j({\bf r})e^{-i\omega_j t}\hat{\alpha}_j+
B^*_j \psi^*_j({\bf r})e^{i\omega_j t}\hat{\alpha}^\dagger_j\right].
\label{eq:quantized-t}
\end{eqnarray}
Using (\ref{eq:quantized-t})  in (\ref{eq:v.1}) immediately gives a 
second relation between the coefficients in (\ref{eq:quantized}), namely
(for $\omega_j > 0$)

\begin{equation}
i\hbar\omega_j B_j=g_0 A_j.
\label{eq:condition2}
\end{equation}
Using (\ref{eq:rho-v.diff}) or (\ref{eq:rho-v.diff_1})
in connection with the expression
(\ref{eq:quantized-t}), we see that the eigenfunctions
$\psi_j({\bf r})$ and eigenvalues $\omega_j$ are determined by

\begin{equation}
-\nabla\cdot\left[{g_0\over m}{\rho}_0({\bf r})\nabla\psi_j({\bf r}) \right]=
\omega^2_j \psi_j({\bf r}).
\label{eq:eigen}
\end{equation}
Combining (\ref{eq:condition2}) with the relation $A^*_j B_j=-i/2$ leads to
(for $\omega_j > 0$)

\begin{equation}
A_j=i\sqrt{\hbar\omega_j\over 2 g_0}\mbox{~~~~~~~~~~~~;~~~~~~~~~~}
B_j=\sqrt{g_0\over 2\hbar\omega_j},
\label{eq:AB}
\end{equation}
where, for convenience, $A_j$ is chosen to be purely imaginary
and $B_j$ is chosen to be real.
Making use of the results (\ref{eq:AB}) in 
(\ref{eq:quantized}), one can now verify after some algebra
that the Hamiltonian (\ref{eq:H.2}) does indeed reduce
to (\ref{eq:H.3}), as assumed.
The harmonic Hamiltonian (\ref{eq:H.3}) describes 
the fluctuations of the condensate $\Phi_0({\bf r})$
in terms of a non-interacting gas of Bose excitations.

Eq.~(\ref{eq:eigen}) has been solved by Stringari \cite{Stringari96-2}
for both isotropic and anisotropic parabolic traps.
For an isotropic harmonic potential,
the solutions are ($j$ represents the usual quantum numbers $(n,\ell,m)$ 
for a spherical potential)
\begin{equation}
\psi_j({\bf r})=c_j P_\ell^{(2n)}({r\over R})\
r^\ell\ Y_{\ell m}(\theta,\phi)\ \Theta(R-r),
\label{eq:eigenfunction}
\end{equation}
where $\Theta(x)$ is the step function.  The associated energy eigenvalues
are found to be \cite{Stringari96-2}
\begin{equation}
\omega_j=\omega_0 \left(2n^2+2n\ell+3n+\ell\right)^{1\over 2}.
\label{eq:eigenvalue}
\end{equation}
In (\ref{eq:eigenfunction}),
$Y_{\ell m}(\theta,\phi)$ are spherical harmonics, while $\displaystyle
P_\ell^{(2n)}(x)=\sum_{m=0}^n \alpha_{2m}(\ell)x^{2m}$
is a polynomial with coefficients satisfying the recurrence relation:
$\alpha_{2m+2}=-\alpha_{2m}(n-m)(2\ell+2m+2n+3)/(m+1)(2\ell+2m+3)$,
with $\alpha_0=1$.  
In the Thomas-Fermi approximation (\ref{eq:rho_0}), the chemical potential
is given by $\mu={1\over 2}m\omega_0^2 R^2$, where
$R$ is the radius at which $\rho_0({\bf r})$ vanishes.
This implies that the integration in (\ref{eq:normalization})
is over a sphere of radius $R$. 
One also finds directly that $\mu=15a\hbar^2 N/2mR^3$, which
can be used to give $R$ in terms of $N$ and $a$. 
One can show that the TF hydrodynamic
approximation should be valid for $\omega_j\alt \mu$ \cite{Stringari96-2}.
The normalization factor 
$c_j$ is determined via (\ref{eq:normalization}),

\begin{equation}
c_j=c_{n\ell m}=\left\{R^{2\ell+3}\int_0^1 dx\ 
x^{2\ell+2}\left[P_\ell^{(2n)}(x) \right]^2\right\}^{-{1\over 2}}.
\label{eq:c_j}
\end{equation}
For $n=0$, one has $c_{o\ell m}=[(2\ell+3)/R^{2\ell+3}]^{1\over 2}$.

For an isotropic parabolic trap, (\ref{eq:eigenfunction}) gives the
normal-mode basis functions which appear in $\delta\hat{\rho}({\bf r},t)$ and
$\delta\hat{\phi}({\bf r},t)$ in (\ref{eq:quantized}).
The normal-mode frequencies (\ref{eq:eigenvalue})
are independent of the interaction strength because
in (\ref{eq:eigen}), we have used the relation
$g_0 \rho_0({\bf r})=\mu-V_{\rm ex}({\bf r})$ given by (\ref{eq:rho_0}).
The corresponding results of
(\ref{eq:eigenfunction}) and (\ref{eq:eigenvalue})
for a uniform Bose fluid \cite{LP80} are $\psi_j({\bf r})\rightarrow
\psi_{\bf k}({\bf r})={1\over \sqrt{V}}e^{i{\bf k}\cdot{\bf r}}$
and $\omega_j\rightarrow\omega ({\bf k})=ck$ with $c^2=g_0 \rho _0/m$.  
In this case, (\ref{eq:rho_0}) gives the usual Bogoliubov
result $\mu=g_0\rho_0$.
We note that only finite-energy solutions of
(\ref{eq:eigen}) are involved in the fluctuations \cite{ERBC96}. 
A zero energy solution
would correspond to the ground state value.
The $n=0$, $\ell=0$ solution of (\ref{eq:eigenfunction}),
 which corresponds to $\psi_j({\bf r})={\rm constant}$ and $\omega_j=0$,
is excluded in all summations such as (\ref{eq:quantized}).
For further discussion of the zero energy solution,
see the end of Section~V.

\section{Green's Functions and Depletion of Condensate}

Writing the quantum field operators
(see Ch.~3 of Ref.~\cite{Griffin93}) in terms of condensate
and non-condensate contributions
$\hat{\psi}({\bf r})\equiv\Phi_0({\bf r})+\tilde{\psi}({\bf r})$,
the diagonal single-particle Green's function $G_{11}$ is given by
[using (\ref{eq:delta_rho})-(\ref{eq:delta_rho.1})]

\begin{eqnarray}
G_{11}({\bf r,r^\prime},t)&\equiv&-\langle\tilde{\psi}^\dagger({\bf r},t)
\tilde{\psi}({\bf r^\prime})\rangle\nonumber\\
&=&-\langle\delta\hat{\Phi}^\dagger({\bf r},t)
\delta\hat{\Phi}({\bf r^\prime})\rangle\nonumber\\
&=&-\Phi_0^*({\bf r})\Phi_0({\bf r^\prime})\biggl[
\langle\delta\hat{A}^\dagger({\bf r},t)
\delta\hat{A}({\bf r^\prime})\rangle+
\langle\delta\hat{\phi}^\dagger({\bf r},t)
\delta\hat{\phi}({\bf r^\prime})\rangle\nonumber\\
&+&i\langle\delta\hat{A}^\dagger({\bf r},t)
\delta\hat{\phi}({\bf r^\prime})\rangle-i\langle
\delta\hat{\phi}^\dagger({\bf r},t)\delta\hat{A}({\bf r^\prime})\rangle\biggr].
\label{eq:G11}
\end{eqnarray}
Using the normal-mode expansion (\ref{eq:quantized}), we find after
some algebra
\begin{eqnarray}
G_{11}({\bf r,r^\prime},t)&=&-\Phi_0^*({\bf r})\Phi_0({\bf r^\prime})
{\sum_{j}} \left\{\left[{\hbar\omega_j\over 8g_0}{1\over \rho_0({\bf r})
\rho_0({\bf r^\prime})}+{g_0\over 2\hbar\omega_j}
\right]C_{j+}({\bf r,r^\prime},t)\right.\nonumber\\
&&+\left.{1\over 4}
\left[{1\over \rho_0({\bf r})}+{1\over \rho_0({\bf r^\prime})} \right]
C_{j-}({\bf r,r^\prime},t)\right\},
\label{eq:G11.1}
\end{eqnarray}
where we have defined
\begin{eqnarray}
C_{j\pm}({\bf r,r^\prime},t)\equiv
\psi_j^*({\bf r})\psi_j({\bf r}^\prime)
N^0(\omega_j)e^{i\omega_j t}\pm
\psi_j({\bf r})\psi_j^*({\bf r}^\prime)
\left[N^0(\omega_j)+1\right]e^{-i\omega_j t}
\label{eq:C+-}
\end{eqnarray}
and $N^0(\omega_j)=[\exp(\beta\hbar\omega_j)-1]^{-1}$ is
the Bose distribution function, with $\beta=1/k_B T$.
In (\ref{eq:G11.1}), the first term involves amplitude fluctuations,
the second term involves phase fluctuations, 
while the third and fourth terms involve the coupling 
between the amplitude and phase fluctuations.
Strictly speaking, this result is only valid at $T\simeq 0$, since we
started with the Gross-Pitaevskii description (\ref{eq:H})
which assumes all the atoms are in the Bose-condensate. The time-ordered 
Green's function equivalent of (\ref{eq:G11.1}) is easily worked out.

In an analogous way, one can calculate $G_{22}$ to be

\begin{eqnarray}
G_{22}({\bf r,r^\prime},t)&\equiv&
-\langle\delta\hat{\Phi}({\bf r},t)
\delta\hat{\Phi}^\dagger({\bf r^\prime})\rangle\nonumber\\
&=&-\Phi_0({\bf r})\Phi_0^*({\bf r^\prime})
{\sum_{j}} \left\{\left[{\hbar\omega_j\over 8g_0}{1\over \rho_0({\bf r})
\rho_0({\bf r^\prime})}+{g_0\over 2\hbar\omega_j}
\right]C_{j+}({\bf r,r^\prime},t)\right.\nonumber\\
&&-\left.{1\over 4}
\left[{1\over \rho_0({\bf r})}+{1\over \rho_0({\bf r^\prime})} \right]
C_{j-}({\bf r,r^\prime},t)\right\},
\label{eq:G22}
\end{eqnarray}
and the off-diagonal single-particle Beliaev Green's functions are

\begin{eqnarray}
G_{12}({\bf r,r^\prime},t)
&\equiv&-\langle\delta\hat{\Phi}^\dagger({\bf r},t)
\delta\hat{\Phi}^\dagger({\bf r^\prime})\rangle\nonumber\\
&=&-\Phi_0^*({\bf r})\Phi_0^*({\bf r^\prime})
{\sum_{j}} \left\{\left[{\hbar\omega_j\over 8g_0}{1\over \rho_0({\bf r})
\rho_0({\bf r^\prime})}-{g_0\over 2\hbar\omega_j}
\right]C_{j+}({\bf r,r^\prime},t)\right.\nonumber\\
&&-\left.{1\over 4}
\left[{1\over \rho_0({\bf r})}-{1\over \rho_0({\bf r^\prime})} \right]
C_{j-}({\bf r,r^\prime},t)\right\}
\label{eq:G12}
\end{eqnarray}
and

\begin{eqnarray}
G_{21}({\bf r,r^\prime},t)
&\equiv&-\langle\delta\hat{\Phi}({\bf r},t)
\delta\hat{\Phi}({\bf r^\prime})\rangle\nonumber\\
&=&-\Phi_0({\bf r})\Phi_0({\bf r^\prime})
{\sum_{j}} \left\{\left[{\hbar\omega_j\over 8g_0}{1\over \rho_0({\bf r})
\rho_0({\bf r^\prime})}-{g_0\over 2\hbar\omega_j}
\right]C_{j+}({\bf r,r^\prime},t)\right.\nonumber\\
&&+\left.{1\over 4}
\left[{1\over \rho_0({\bf r})}-{1\over \rho_0({\bf r^\prime})} \right]
C_{j-}({\bf r,r^\prime},t)\right\}
\label{eq:G21}
\end{eqnarray}

As an application of the above results, we calculate the
depletion of the condensate density near $T=0$. Using (\ref{eq:G11.1})
and (\ref{eq:C+-}), one obtains

\begin{eqnarray}
\langle\tilde{\rho}({\bf r})\rangle&\equiv&-G_{11}({\bf r,r},t=0) \nonumber\\
&=&{\sum_{j}} |\psi_j({\bf r})|^2\ {1\over 2}
\left[{\hbar\omega_j\over 4g_0 \rho_0({\bf r})}
+{g_0\rho_0({\bf r})\over \hbar\omega_j}-1\right]~~~~~~~~~~~~~~(T\simeq 0).
\label{eq:depletion}
\end{eqnarray}
For a spatially uniform Bose-condensed gas ($\rho_0({\bf r})={\rm constant}$ 
and $\omega_j\rightarrow\omega ({\bf k})=ck$), 
the phase fluctuations (second term) are the dominant contribution
to the expression in (\ref{eq:depletion}). It then reduces to \cite{LP80}
\begin{equation}
\tilde{\rho}=\rho-\rho_0={1\over 2}\int{d{\bf k}\over (2\pi)^3}
{g_0\rho_0\over \hbar\omega({\bf k})}.
\label{eq:depletion.uniform}
\end{equation}
In contrast, the phase fluctuations are not so dominant
in a trapped Bose gas because their energy is finite.
We note that (\ref{eq:depletion}) is similar to recent results
based on a variational solution of the full 
Bogoliubov equations \cite{Fetter96-2}.  

\begin{figure}
\postscript{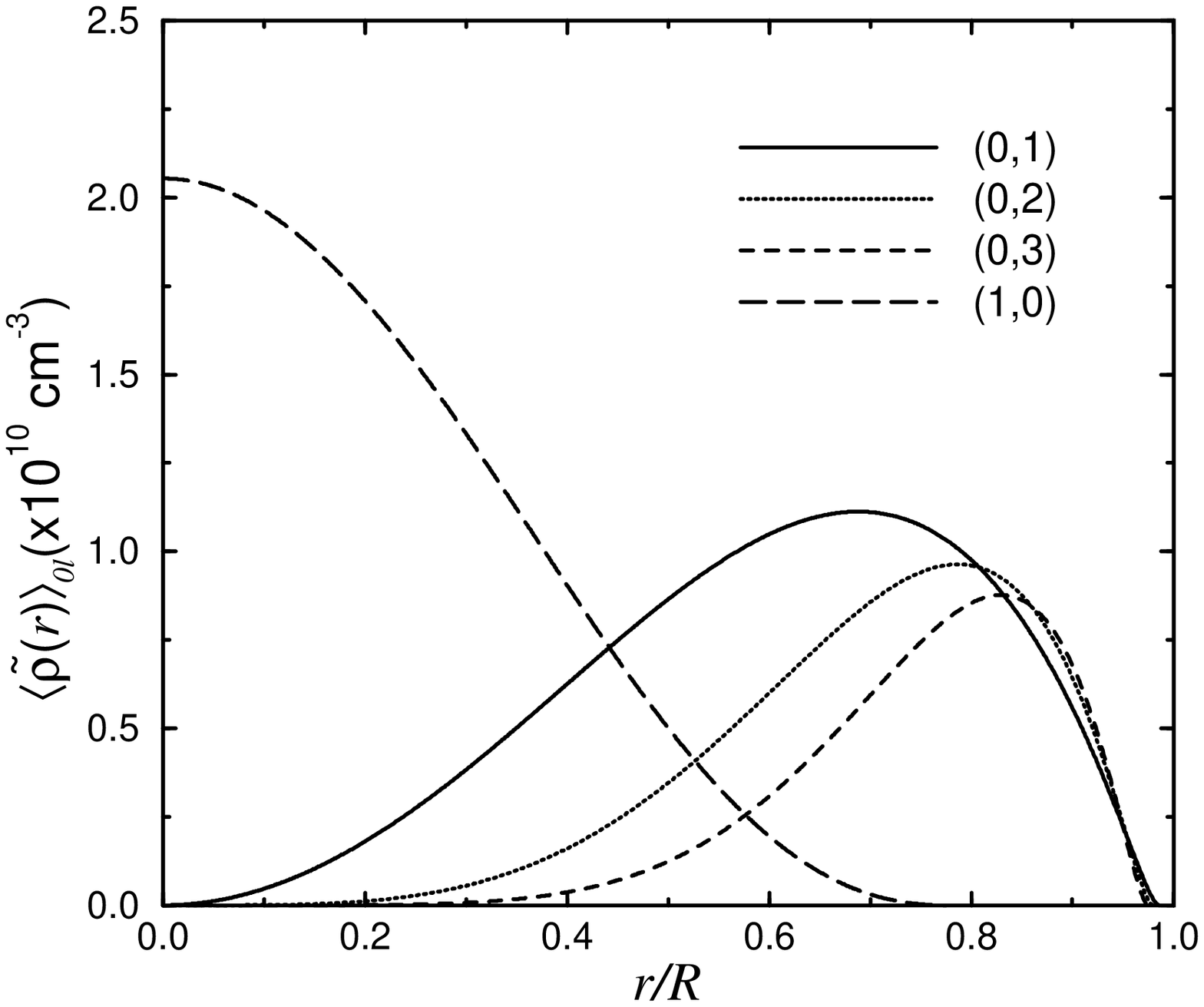}
\caption{Plot of the depletion given by (\protect\ref{eq:depletion})
of the local condensate density
due to the contributions of several low-energy excitations
specified by $(n,\ell)$ \protect\cite{statement_paper6}.
The parameters used are: $a_{HO}=10^{4}{\rm \AA}$,
$a=50{\rm \AA}$,  and $N=10^5$, which give 
$R=6\times 10^{4}{\rm \AA}$.}
\label{fig1}
\end{figure}

As discussed by Stringari \cite{Stringari96-2},
the collective mode energies in (\ref{eq:eigenvalue})
in the $n=0$ case reduce to the simple
expression $\omega_j\rightarrow\omega_{0\ell}=\omega_0 \sqrt{\ell}$,
for $\ell$ finite.
The $\ell$-channel contribution in (\ref{eq:depletion}) for these $n=0$
``surface modes'' \cite{Fetter96-1} is given by

\begin{eqnarray}
\langle\tilde{\rho}({\bf r})\rangle_{0,\ell}=
{1\over 16\pi R^3}(2\ell+1)(2\ell+3)\ \bar{r}^{2\ell}\left(
{\sqrt{\ell}\ \bar{a}_{HO}^2\over 1-\bar{r}^2}+
{1-\bar{r}^2\over \sqrt{\ell}\ \bar{a}_{HO}^2}-2\right),
\label{eq:depletion.1}
\end{eqnarray}
where $\sum_{m}|Y_{\ell m}(\theta,\phi)|^2=(2\ell+1)/4\pi$ has
been used and we have defined
$\bar{r}\equiv r/R$, $\bar{a}_{HO}\equiv a_{HO}/R$.
The $n=1$, $\ell=0$ contribution to the depletion is ($\omega_{1,0}=
\sqrt{5}\omega_0$)

\begin{eqnarray}
\langle\tilde{\rho}({\bf r})\rangle_{1,0}=
{63\over 64\pi R^3}\left(1-{5\over 3}\bar{r}^{2}\right)^2
\left({\sqrt{5}\ \bar{a}_{HO}^2\over 1-\bar{r}^2}+
{1-\bar{r}^2\over \sqrt{5}\ \bar{a}_{HO}^2}-2\right).
\label{eq:depletion.2}
\end{eqnarray}
For illustration, in Fig.~\ref{fig1} we show the depletion
$\langle\tilde{\rho}({\bf r})\rangle$ as a function of $r$
from the lowest-energy modes described by (\ref{eq:depletion.1}) 
and (\ref{eq:depletion.2}) \cite{statement_paper6}.
We have assumed $N=10^5$ trapped atoms, an {\it s}-wave scattering length
$a=50{\rm \AA}$ and $a_{HO}=10^{4}{\rm \AA}$.
These parameters give $R=6\times 10^{4}{\rm \AA}$ \cite{DPS96}. 
The local condensate density is
given by (\ref{eq:rho_0}), namely

\begin{equation}
\rho_0({\bf r})={R^2\over 8\pi a a_{HO}^4} \left(1-\bar{r}^2\right)
\Theta(1-\bar{r}).
\label{eq:rho_0.explicit}
\end{equation}
For the parameters used in Fig.~\ref{fig1}, we find 
$\rho_0({\bf r})=(1-\bar{r}^2)\ 2.8\times 10^{14}{\rm cm}^{-3}$ 
and thus the $T=0$ condensate depletion is very small, as expected.
For these parameters, we note that 
$\mu={1\over 2}\left({R\over a_{HO}}\right)^2
\hbar\omega_0=18\hbar\omega_0$ and 
hence these frequencies satisfy the condition
$\omega\alt \mu$ for which the TF approximation is valid \cite{Stringari96-2}.

\section{Dynamic Structure Factor}

We next calculate the light scattering cross section which
is proportional to the dynamic structure 
factor $S({\bf q},\omega)$. The latter is given by the Fourier
transform of the density-density correlation function.
We recall that the density can be split into contribution
involving atoms being excited in/out of the condensate and 
contribution from non-condensate atoms. Thus we have  
(see Ch.~3 of \cite{Griffin93})

\begin{eqnarray}
\hat{\rho}({\bf r})&\equiv& \hat{\psi}^\dagger({\bf r})
\hat{\psi}({\bf r})\nonumber\\
&=& |\Phi_0({\bf r})|^2+[\Phi_0^*({\bf r})\tilde{\psi}({\bf r})+
\Phi_0({\bf r})\tilde{\psi}^\dagger({\bf r})]
+\tilde{\psi}^\dagger({\bf r})\tilde{\psi}({\bf r})\nonumber\\
&\equiv& \rho_0({\bf r})+\delta\hat{\rho}_c({\bf r})+\tilde{\rho}({\bf r}),
\label{eq:rho_expansion}
\end{eqnarray}
where $\tilde{\rho}({\bf r})\equiv \tilde{\psi}^\dagger({\bf r})
\tilde{\psi}({\bf r})$ is the non-condensate local density operator.
In the present discussion of a dilute Bose gas, it will be
assumed that one can
neglect $\tilde{\rho}({\bf r})$ as being small at $T=0$.
The density-density correlation function
due to the condensate density fluctuations is easily calculated to be 
[using (\ref{eq:delta_rho.1}), (\ref{eq:G11}), and (\ref{eq:G11.1})]
 
\begin{eqnarray}
\langle 
\delta\hat{\rho}_c({\bf r},t)\delta\hat{\rho}_c({\bf r^\prime})\rangle
&=&4\rho_0({\bf r})\rho_0({\bf r}^\prime)
\langle\delta\hat{A}({\bf r},t)
\delta\hat{A}({\bf r^\prime})\rangle\nonumber\\
&=&{\sum_{j}}  {\hbar\omega_j\over 2g_0}C_{j+}({\bf r,r^\prime},t),
\label{eq:chi}
\end{eqnarray}
where the function $C_{j+}({\bf r,r^\prime},t)$ has been 
defined in (\ref{eq:C+-}). 
Thus for the problem being considered, the condensate contribution
to $S({\bf q},\omega)$ is given by  

\begin{eqnarray}
S_c({\bf q},\omega) &\propto&
\int d{\bf r}e^{-i{\bf q\cdot r}}\int d{\bf r^\prime}
e^{i{\bf q\cdot r^\prime}}\int_{-\infty}^{\infty} dt e^{i\omega t} \langle  
\delta\hat{\rho}_c({\bf r},t)\delta\hat{\rho}_c({\bf r^\prime})\rangle
\nonumber\\
&=&{\sum_{j}}  {\hbar\omega_j\over 2g_0}|\psi_j({\bf q})|^2
\left\{\left[N^0(\omega_j)+1\right]\delta(\omega-\omega_j)
+N^0(\omega_j)\delta(\omega+\omega_j)\right\},
\label{eq:chi.1}
\end{eqnarray} 
where

\begin{equation}
\psi_j({\bf q})=\int\ d{\bf r} e^{-i{\bf q}\cdot{\bf r}}
\psi_j({\bf r}).
\label{eq:FT}
\end{equation}
This is for a momentum transfer $\hbar{\bf q}$ and energy
transfer $\hbar\omega$ given to the Bose condensate.

Since we are only dealing with condensate fluctuations
(density fluctuations involving atoms entering/leaving the 
condensate), it is trivial that the poles of $S_c({\bf q},\omega)$
in (\ref{eq:chi.1}) are identical to those of the single-particle
Green's functions in (\ref{eq:G11}) and (\ref{eq:G12}).
However, this sharing of poles is a general feature of Bose-condensed 
fluids even when one includes non-condensate contributions
(see Ch.~5 of Ref.~\cite{Griffin93}).
While (\ref{eq:chi.1}) is based on a calculation which is only 
valid at $T\simeq 0$, we expect that 
more generally, it will approximately describe the contribution related
to exciting excitations out of the condensate. In this context,
it is useful to give the two limiting expressions

\begin{eqnarray}
S_c({\bf q},\omega)\rightarrow
\left\{
\begin{array}{lr}
\displaystyle {\sum_{j}}  {\hbar\omega_j\over 2g_0}|\psi_j({\bf q})|^2 
\delta(\omega-\omega_j) & \hspace{0.5cm}
\makebox[3cm]{{\rm if} $k_B T\ll \hbar\omega_j$} \\ 
\displaystyle {\sum_{j}}  {k_B T\over 2g_0}|\psi_j({\bf q})|^2 
[\delta(\omega-\omega_j)+\delta(\omega+\omega_j)] & \hspace{0.5cm}
\makebox[3cm]{{\rm if} $k_B T\gg \hbar\omega_j$} \\ 
\end{array}
\right..
\label{eq:chi.2}
\end{eqnarray}
Of course, (\ref{eq:chi.1}) and (\ref{eq:chi.2})
do not include the scattering from thermally excited atoms present at $T\neq 0$
or the two-excitation (``multiphonon'') contributions \cite{Griffin93,GW96}.
These are briefly discussed at the end of this section.

In the analogous calculation of $S_c({\bf q},\omega)$ 
for a weakly interacting 
{\em homogeneous} Bose-condensed gas \cite{Griffin93,GW96}, one 
picks up a {\em single} sharp peak at the 
quasiparticle energy $\omega({\bf q})$
corresponding to the momentum transfer involved.  In a trapped Bose gas, 
in contrast, $S_c({\bf q},\omega)$ in (\ref{eq:chi.1}) is 
seen to be a weighted sum
of {\em all} the normal modes describing the fluctuating inhomogeneous
condensate,  of frequency $\omega_j$ and with a weight 
proportional to $\omega_j|\psi_j({\bf q})|^2$.

Using the well-known expansion (for ${\bf q}$ parallel to $z$-axis)

\begin{equation}
e^{-i{\bf q\cdot r}}=\sum^\infty_{\ell=0}
\sqrt{4\pi (2\ell+1)}(-i)^\ell j_\ell(qr)Y_{\ell 0}(\theta),
\label{eq:expansion}
\end{equation}
where $j_\ell$ is the $\ell$-order Bessel function, 
one can use (\ref{eq:eigenfunction}) to directly obtain the Fourier
transform of the eigenfunction $\psi_j({\bf r})$, namely
\begin{equation}
\psi_{n\ell m}({\bf q})=c_{n\ell m}(-i)^\ell\delta_{m,0}\sqrt{4\pi (2\ell+1)}\ 
R^{\ell+3}\int_0^1 dx\ x^{\ell+2}P_\ell^{(2n)}(x)j_\ell(qRx),
\label{eq:psi_q}
\end{equation}
where the normalization factor $c_{n\ell m}$ is given by (\ref{eq:c_j}). 
We note that as a result of (\ref{eq:expansion}), only the states
$\psi_j({\bf r})$ with $m=0$ have a finite weight in
$S_c({\bf q},\omega)$ in (\ref{eq:chi.1}).

\begin{figure}
\postscript{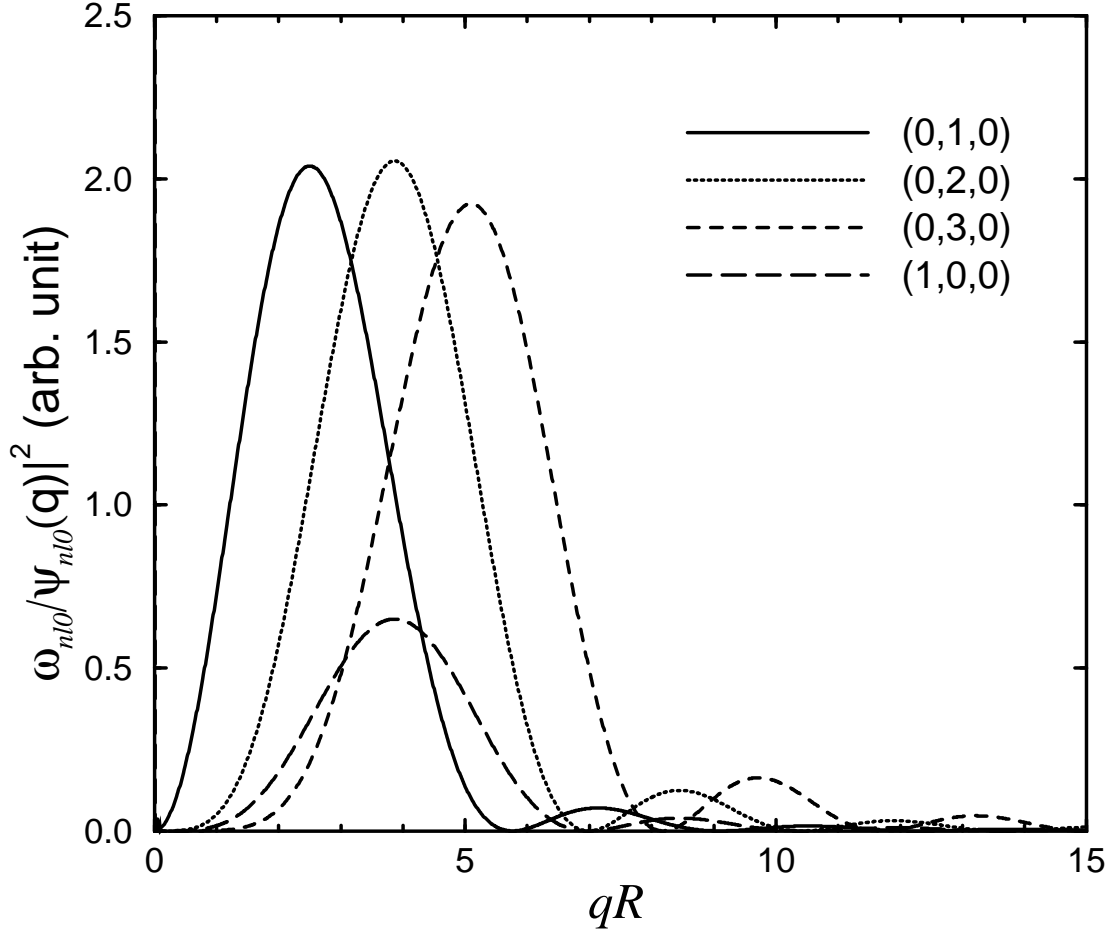}
\caption{Plot of $\omega_j|\psi_j({\bf q})|^2$ given in 
(\protect\ref{eq:chi.2}) as a function of dimensionless wavevector $qR$, 
where $R$ is the size of the condensate. These functions determine the weight 
of the light-scattering cross section in (\protect\ref{eq:chi.2})
of the corresponding collective modes of energy $\omega_{n\ell 0}$ 
(see Figs.~\protect\ref{fig3} and \protect\ref{fig4}).}
\label{fig2}
\end{figure}

In Fig.~\ref{fig2}, we have plotted 
$\omega_j|\psi_j({\bf q})|^2$ which appears in the formula 
(\ref{eq:chi.1}) for
$S_c({\bf q},\omega)$ as a function of the dimensionless 
wavevector $qR\ $ for excitations
$(n,\ell)=(0,1), ~(0,2), ~(0,3)$ and $(1,0)$, with
$m=0$ in all cases. 
As one can see from Fig.~\ref{fig2}, the strongest weights for
these collective modes appear for $qR\alt 6$. 
As an example, for the parameters used in Fig.~\ref{fig1}
to give $R=6\times 10^{-4}{\rm cm}$, this means that
ideally, the momentum transfer $q$ in a light-scattering experiment
should be $q\alt 10^{4}{\rm cm}^{-1}$
in order to pick up the strong spectral weight from the low-energy
collective modes. 
For $qR\agt 1$, the density fluctuations being
probed have a wavelength smaller than the 
size of the condensate $2R$. The Thomas-Fermi approximation 
(\ref{eq:rho_0}) we have
used should be adequate. In contrast, the TF approximation is 
probably not very good for $qR\ll 1$ since
the results could be sensitive to the abrupt vanishing of $\rho_0({\bf r})$
for $r>R$.

\begin{figure}
\postscript{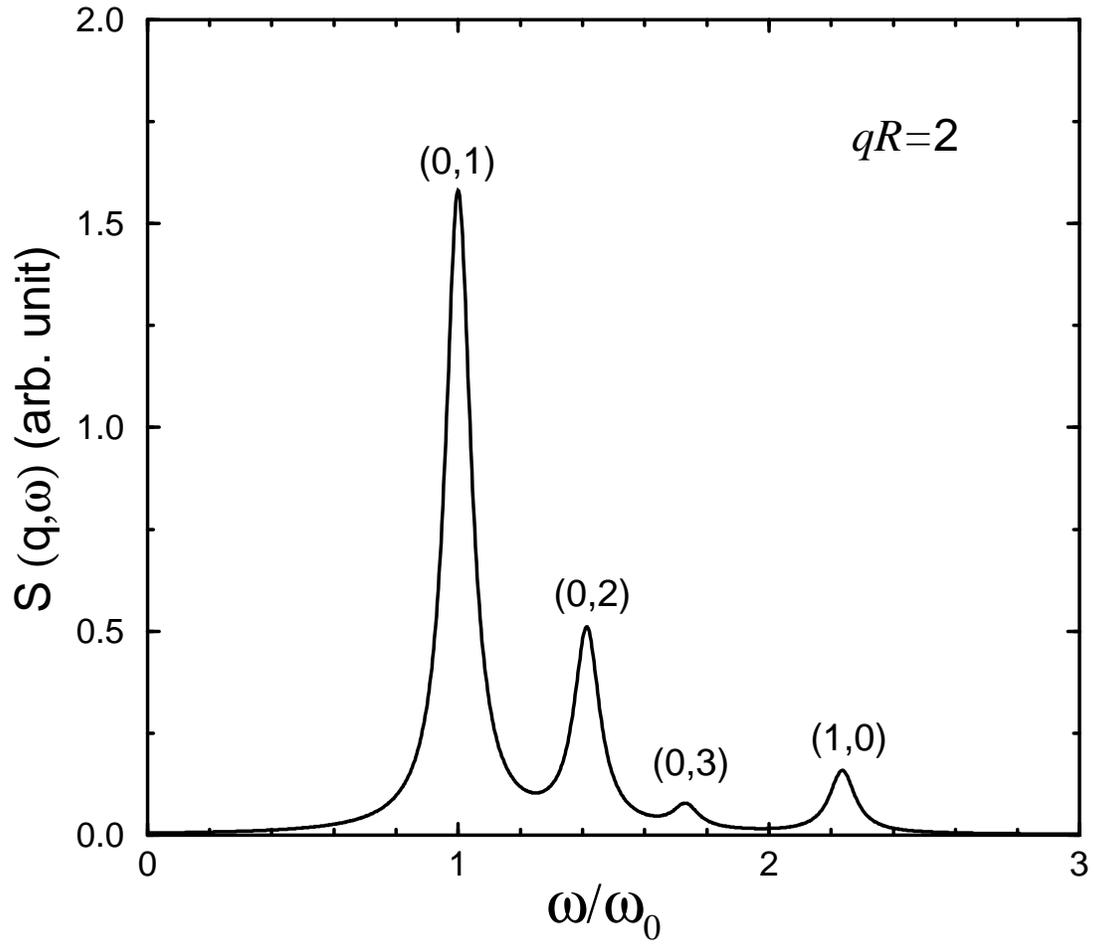}
\caption{Plot of the $T=0$ dynamic structure factor
$S({\bf q},\omega)$ {\rm vs} $\omega$,
using a Lorentzian broadening of the delta function peaks
($\Gamma=0.05\omega_0$) and a momentum transfer $qR=2$.
Only the contributions from the low-frequency modes $(n,\ell,0)$ are included.}
\label{fig3}
\end{figure}

\begin{figure}
\postscript{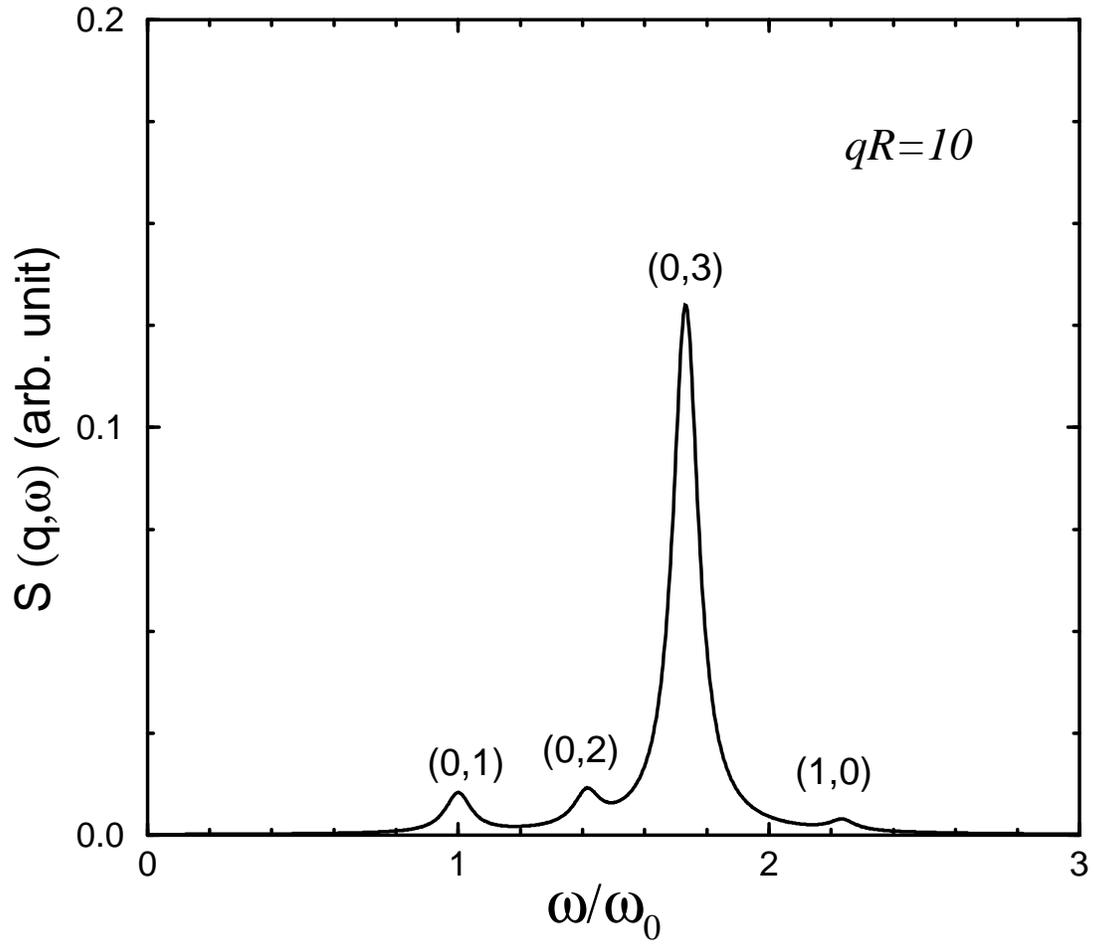}
\caption{Same plot as Fig.~\protect\ref{fig3}, for $qR=10$.
Note the much smaller intensity compared to the results for
$qR=2$ in Fig.~\protect\ref{fig3}.}
\label{fig4}
\end{figure}

In  Fig.~\ref{fig3}, we plot the $T=0$ dynamic structure factor
$S_c({\bf q},\omega)$ given by (\protect\ref{eq:chi.2}).
For clarity and as a description of finite
energy resolution, we have broadened the delta function
$\delta(\omega)$ using a Lorentzian, with a width of $\Gamma=0.05\omega_0$.
The momentum transfer corresponds to $qR=2$. In Fig.~\ref{fig4}, we give 
results for a  higher momentum transfer $qR=10$ using the same 
arbitrary units as in Fig.~\ref{fig3}, showing 
a much weaker light-scattering intensity.

Besides the condensate contribution to
$S({\bf q},\omega)$ given by
(\ref{eq:chi.1}), we can use our
hydrodynamic approximation to estimate the contribution
of the non-condensate part.
Using (\ref{eq:rho_expansion}),
the density-density correlation function arising from the
non-condensate fluctuations can be expressed in terms of the
single-particle Green's function $G_{\alpha\beta}$ discussed in Section~III,

\begin{eqnarray}
&&\langle \tilde{\rho}({\bf r},t)\tilde{\rho}({\bf r^\prime})\rangle
= \langle \delta\hat{\Phi}^\dagger({\bf r},t)
\delta\hat{\Phi}({\bf r},t)\delta\hat{\Phi}^\dagger({\bf r}^\prime) 
\delta\hat{\Phi}({\bf r}^\prime)\rangle\nonumber\\
&&=\langle \delta\hat{\Phi}^\dagger({\bf r},t)
\delta\hat{\Phi}({\bf r}^\prime)\rangle
\langle\delta\hat{\Phi}({\bf r},t)
\delta\hat{\Phi}^\dagger({\bf r}^\prime)\rangle
+\langle\delta\hat{\Phi}({\bf r},t)
\delta\hat{\Phi}({\bf r}^\prime)\rangle
\langle \delta\hat{\Phi}^\dagger({\bf r},t)
\delta\hat{\Phi}^\dagger({\bf r}^\prime)\rangle\nonumber\\
&&=G_{11}({\bf r,r^\prime},t)G_{22}({\bf r,r^\prime},t)+
G_{21}({\bf r,r^\prime},t)G_{12}({\bf r,r^\prime},t).
\label{eq:rho_tilde-rho_tilde}
\end{eqnarray}
After some algebra, using (\ref{eq:G11.1})-(\ref{eq:G21}),
we obtain the $T=0$ contribution due to creating two excitations,

\begin{eqnarray}
\tilde{S}({\bf q},\omega) &\propto&
\int d{\bf r}e^{-i{\bf q\cdot r}}\int d{\bf r^\prime}
e^{i{\bf q\cdot r^\prime}}\int_{-\infty}^{\infty} dt e^{i\omega t}
\langle \tilde{\rho}({\bf r},t)\tilde{\rho}({\bf r^\prime})\rangle
\nonumber\\
&=&{\sum_{i,j}}\ 2\ |B_{ij}({\bf q})|^2\delta(\omega-\omega_i-\omega_j),
\label{eq:chi.non-condensate}
\end{eqnarray}
where we have defined

\begin{equation}
B_{ij}({\bf q})\equiv \int d{\bf r} e^{-i{\bf q}\cdot{\bf r}}
\left[{g_0\rho_0({\bf r})\over 2\hbar\sqrt{\omega_i \omega_j}}-
{\hbar\sqrt{\omega_i\omega_j}\over 8g_0 \rho_0({\bf r})} \right]
\psi_i({\bf r})\psi_j({\bf r}).
\label{eq:FT.B}
\end{equation}
This result is the generalization of a similar calculation
for a uniform Bose gas given in Ref.~\cite{Griffin93}.
Of course, at $T=0$, this contribution is much smaller than
the condensate contribution $S_c({\bf q},\omega)$ in (\ref{eq:chi.1})
since most of the atoms are in the condensate, {\em i.e.},
$\tilde{\rho}\ll \rho_c$.
Using (\ref{eq:eigenvalue}), one can see that $\tilde{S}({\bf q},\omega)$ 
contributes only when $\omega\geq 2\omega_0$.

The general expressions (\ref{eq:chi.1}) and (\ref{eq:chi.non-condensate})
given in this section are also valid for
anisotropic parabolic wells. Stringari \cite{Stringari96-2}
has discussed the solutions
$\psi_j({\bf r})$ of (\ref{eq:eigen}) for traps with
axial symmetry about the $z$-axis. In this case, 
the azimuthal quantum number $m$ still characterizes the normal-mode 
eigenvectors $\psi_j({\bf r})$.

\section{Relation to Bogoliubov coupled equations}

The standard approach to study the dynamics of a $T=0$ Bose condensate
(uniform or non-uniform) is based on the Bogoliubov approximation
\cite{Fetter72,Griffin96}.
The non-condensate field operators are decomposed into 
positive- and negative-energy Bose fluctuations

\begin{eqnarray}
\tilde{\psi}({\bf r},t)&=&e^{i\phi_0({\bf r})}
{\sum_{j}} \left[u_j ({\bf r})e^{-iE_j t/\hbar}
\hat{\alpha}_j-v_j^* ({\bf r})e^{iE_j t/\hbar} \hat{\alpha}^\dagger_j\right]
\nonumber\\
\tilde{\psi}^\dagger({\bf r},t)&=&e^{-i\phi_0({\bf r})}
{\sum_{j}} \left[u_j^* ({\bf r})e^{iE_j t/\hbar}
\hat{\alpha}^\dagger_j-v_j({\bf r})e^{-iE_j t/\hbar} \hat{\alpha}_j\right],
\label{eq:HFB}
\end{eqnarray}
where $\hat{\alpha}_j$, $\hat{\alpha}^\dagger_j$
satisfy Bose commutation relations.
The amplitude functions $u_j ({\bf r})$ and $v_j ({\bf r})$ in (\ref{eq:HFB})
are given by the solutions of the coupled eigenvalue equations 
\cite{Fetter72}

\begin{eqnarray}
\hat{L}u_j ({\bf r})-g_0|\Phi_0({\bf r})|^2 v_j({\bf r})&=&E_j u_j ({\bf r})
\nonumber\\
\hat{L}^* v_j ({\bf r})-g_0|\Phi_0({\bf r})|^2 u_j({\bf r})&=&-E_j v_j ({\bf r})
\label{eq:HFB.coupled}
\end{eqnarray}
with 
\begin{equation}
\hat{L}\equiv {1\over 2m}\left[-i\hbar\nabla+m{\bf v}_0({\bf r})\right]^2
-\mu+V_{\rm ex}
({\bf r})+2g_0|\Phi_0({\bf r})|^2.
\label{eq:L.def.phase}
\end{equation}
As noted by Fetter \cite{Fetter96-2}, 
the solutions of (\ref{eq:HFB.coupled}) for $u_j$ and $v_j$ are basically
linear combinations of $\delta\rho$ and $\delta\phi$ in the TF
approximation studied 
in this paper. Within this approximation and setting ${\bf v}_0({\bf r})=0$,
(\ref{eq:delta_Phi}) and (\ref{eq:delta_rho.1}) give

\begin{equation}
\tilde{\psi}({\bf r},t)=\delta\hat{\Phi}({\bf r},t)=
\Phi_0({\bf r})\left[{\delta\hat{\rho}({\bf r},t)\over 2\rho_0({\bf r})}+
i\delta\hat{\phi}({\bf r},t)\right].
\label{eq:tildephi_com}
\end{equation}
Comparing terms in (\ref{eq:HFB}) with (\ref{eq:quantized}),
(\ref{eq:tildephi_com}) immediately gives 

\begin{eqnarray}
u_j ({\bf r})&=&i\left(\sqrt{g_0\rho_0({\bf r})\over 2\hbar \omega_j}+
\sqrt{\hbar \omega_j\over 8g_0\rho_0({\bf r})}\ \right)\psi_j({\bf r})
\nonumber\\
v_j ({\bf r})&=&i\left(\sqrt{g_0\rho_0({\bf r})\over 2\hbar \omega_j}-
\sqrt{\hbar \omega_j\over 8g_0\rho_0({\bf r})}\ \right)\psi_j({\bf r}),
\label{eq:uv}
\end{eqnarray}
with $E_j=\hbar \omega_j$. Here we have used 
$|\Phi_0({\bf r})|=\sqrt{\rho_0({\bf r})}$.
One can easily verify using (\ref{eq:normalization}) and 
(\ref{eq:completeness}) that $u_j({\bf r})$ and $v_j({\bf r})$
given by (\ref{eq:uv}) satisfy the usual orthogonality and
completeness relations.

For completeness, it is useful to show explicitly how
the Bogoliubov equations
(\ref{eq:HFB.coupled}) can be used 
to derive the key equations (\ref{eq:rho.1}) and (\ref{eq:v.1}) 
on which the TF hydrodynamic description is based.
Using (\ref{eq:rho_0}) , the operator in (\ref{eq:L.def.phase}) can
be written as $\hat{L}=\hat{L}_{GP}+g_0|\Phi_0({\bf r})|^2$, where

\begin{equation}
\hat{L}_{GP}\equiv -{\hbar^2\nabla^2\over 2m}-\mu+V_{\rm ex}({\bf r})
+g_0|\Phi_0({\bf r})|^2.
\label{eq:L.def.phase.approxi}
\end{equation}
Subtracting and adding the two equations in (\ref{eq:HFB.coupled}) gives

\begin{eqnarray}
&\displaystyle \hat{L}_{GP} 
\left[u_j({\bf r })-v_j({\bf r })\right]+
2g_0\rho_0({\bf r})\left[u_j({\bf r })-v_j({\bf r })\right]
=E_j\left[u_j({\bf r })+v_j({\bf r })\right]&
\label{eq:L1-L2}\\
&\displaystyle \hat{L}_{GP} 
\left[u_j({\bf r })+v_j({\bf r })\right]
=E_j\left[u_j({\bf r })-v_j({\bf r })\right].&
\label{eq:L1+L2}
\end{eqnarray}
Making use of (\ref{eq:uv}), (\ref{eq:quantized}), and (\ref{eq:AB}), 
one finds by direct comparison

\begin{eqnarray}
u_j({\bf r })-v_j({\bf r })&=&i\sqrt{\hbar \omega_j\over 
2g_0\rho_0({\bf r})}\ \psi_j({\bf r})\equiv
{\delta\rho_j({\bf r})\over |\Phi_0({\bf r})|}
\nonumber\\
u_j({\bf r })+v_j({\bf r })&=&2i\sqrt{g_0\rho_0({\bf r})\over 
2\hbar \omega_j}\
\psi_j({\bf r})\equiv 2i|\Phi_0({\bf r})|\delta\phi_j({\bf r}),
\label{eq:uv.quantized}
\end{eqnarray}
where, using (\ref{eq:quantized}) and (\ref{eq:AB}), we have defined 
$\delta\rho_j({\bf r})\equiv i\sqrt{\hbar \omega_j\over 2g_0}\psi_j({\bf r})$ 
and $\delta\phi_j({\bf r})\equiv \sqrt{g_0\over 2\hbar \omega_j} 
\psi_j({\bf r})$.
Taking into account (\ref{eq:rho_0}), we see that the 
first term on the r.h.s. of (\ref{eq:L1-L2}) reduces to
$-{\hbar^2\over 2m}\nabla^2[\delta\rho_j({\bf r})/
|\Phi_0({\bf r})|]$. Since this is omitted in the TFA [see (\ref{eq:neglect})],
(\ref{eq:L1-L2}) gives

\begin{equation}
iE_j \delta\phi_j({\bf r})=g_0 \delta\rho_j({\bf r}).
\label{eq:key.2_Bogo}
\end{equation}
Using this in (\ref{eq:quantized-t}), this is equivalent to (\ref{eq:v.1}) or

\begin{equation}
\hbar{\partial\delta\hat{\phi}({\bf r},t)\over \partial t}=
-g_0\delta\hat{\rho}({\bf r},t).
\label{eq:key.2_final}
\end{equation}
Similarly, substituting (\ref{eq:uv.quantized}) into (\ref{eq:L1+L2}) gives

\begin{equation}
{\hbar\over m}|\Phi_0({\bf r})|
\nabla^2\left[|\Phi_0({\bf r})|\delta\phi_j({\bf r})\right]
=i{E_j\over \hbar} \delta\rho_j({\bf r}).
\label{eq:key.1_Bogo}
\end{equation}
Within the TF approximation ({\em i.e.},
neglecting terms involving $\nabla^2|\Phi_0({\bf r})|$),
(\ref{eq:key.1_Bogo}) is equivalent to

\begin{equation}
\nabla\cdot\left[|\Phi_0({\bf r})|^2\delta{\bf v}_j({\bf r})\right]
=i{E_j\over \hbar} \delta\rho_j({\bf r}),
\label{eq:key.1_Bogo.1}
\end{equation}
which, using (\ref{eq:quantized-t}), is equivalent to (\ref{eq:rho.1}).

We note that while (\ref{eq:L1-L2}) and (\ref{eq:L1+L2}) 
describe fluctuations of the equilibrium condensate,
formally they also exhibit a special solution
(labelled by $s$) corresponding to $u_s({\bf r})=v_s({\bf r})$
with zero energy $E_s=0$ \cite{Fetter72,Griffin96}. 
If one identifies $u_s({\bf r})$
with $|\Phi({\bf r})|$, (\ref{eq:L1+L2}) is then equivalent to the static GP
equation in (\ref{eq:key.2_static}). 
This zero-energy solution still appears in
the TF hydrodynamic approximation described by (\ref{eq:uv.quantized}),
in which case we see that $\delta\rho_s({\bf r})=0$ and
$u_s({\bf r})=i|\Phi_0({\bf r})|\delta \phi_s({\bf r})$.
This is a zero energy solution of (\ref{eq:rho-v.diff_1}) 
corresponding 
to a phase change $\delta\phi_s({\bf r})$ which is independent of ${\bf r}$. 
This zero frequency fluctuation of 
the order parameter has a very simple physical meaning,
namely it involves uniform (rigid) phase change $\delta\phi_s$ at all 
points, but with no associated amplitude (or density) fluctuation 
($\delta\rho_s({\bf r})=0$). This is the expected zero frequency 
mode in a system with a Bose broken symmetry.

\section{Concluding Remarks}

The quantized expressions of density and phase fluctuations give a
simple way to calculate the contribution of low-energy
condensate fluctuations to various observable quantities. The
results given in this paper are based on the ``hydrodynamic description''
in terms of the amplitude and phase of the condensate wavefunction
at zero temperature, as formulated by Stringari \cite{Stringari96-2}
for trapped Bose fluids, in which the kinetic energy contribution
due to density fluctuations is neglected. 
While this Thomas-Fermi hydrodynamic description is only adequate for
the low-frequency modes ($\omega_j \alt \mu$),
it is much simpler than dealing with the full set of
coupled Bogoliubov equations given by (\ref{eq:HFB.coupled}). 

Calculations based on these normal modes allow one to exhibit the 
role of amplitude and phase fluctuations of the condensate in a very 
explicit fashion, complementing a more complete calculation
based on numerical solutions of the 
Bogoliubov equations \cite{ERBC96,Fetter96-1,ERBDC96}.
We have illustrated this formalism by evaluating the
single-particle diagonal
and off-diagonal Green's functions, the $T=0$ condensate depletion,
and the dynamic structure factor $S({\bf q},\omega)$, 
taking into account the contribution of the lowest-frequency modes.
These specific calculations are carried out for an isotropic parabolic
trap but can be generalized for an anisotropic trap using the results in
Ref.~\cite{Stringari96-2}.

\acknowledgements
We thank Sandro Stringari for discussion and useful suggestions.
This work was supported by a research grant from NSERC of Canada.



\end{document}